\documentclass[12pt]{article}

\usepackage{scicite}
 \usepackage{graphicx}
 \usepackage{dcolumn}
 \usepackage{bm}
 \usepackage{pstricks}
 \usepackage{float}
 \usepackage{epsfig}
 \usepackage{caption}
\usepackage{subcaption}
 \usepackage{multirow} 
 \usepackage{amsmath}
 \usepackage{amssymb}
 \usepackage{amsfonts}
 
 \usepackage{amssymb}
 \usepackage{lineno, blindtext}
 \usepackage{color}
\usepackage{epsfig}

\newenvironment{sciabstract}{%
\begin{quote} \bf}
{\end{quote}}

\title{Concentrated Radiative Cooling}

\author
{Joseph Peoples,$^{1}$ Yu-Wei Hung,$^{2}$ Xiangyu Li,$^{1}$\\
Daniel Gallagher,$^{1}$ Nathan Fruehe,$^{1}$ Anil Yuksel,$^{3}$\\
 James Braun,$^{1}$ Travis Horton,$^{2}$
 Xiulin Ruan$^{1\ast}$\\
\\
\normalsize{$^{1}$School of Mechanical Engineering, Purdue University, West Lafayette, IN 47907, USA,}\\
\normalsize{$^{2}$School of Civil Engineering, Purdue University, West Lafayette, IN 47907, USA,}\\
\normalsize{$^{3}$IBM Corporation, Austin, TX, 78758, USA}\\
\\
\normalsize{$^\ast$To whom correspondence should be addressed; E-mail:  ruan@purdue.edu.}
}

\date{}

\begin{document} 


\baselineskip24pt


\maketitle


\begin{sciabstract}
A fundamental limit of current radiative cooling systems is that only the top surface facing deep-space can provide the radiative cooling effect, while the bottom surface cannot.  Here, we propose and experimentally demonstrate a concept of ``concentrated radiative cooling" by nesting a radiative cooling system in a mid-infrared reflective trough, so that the lower surface, which does not contribute to radiative cooling in previous systems, can radiate heat to deep-space via the reflective trough. Field experiments show that the temperature drop of a radiative cooling pipe with the trough is more than double that of the standalone radiative cooling pipe. Furthermore, by integrating the concentrated radiative cooling system as a preconditioner in an air conditioning system, we predict electricity savings of $>75\%$ in Phoenix, AZ, and $>80\%$ in Reno, NV, for a single-story commercial building.
\end{sciabstract}


\section*{Introduction}

In 2018, 9.3\% of the total generated electricity in the United States went to space cooling and refrigeration of commercial buildings, equating to 164.7 million metric tons of carbon dioxide emission, just for commercial cooling \cite{U.S.EIA2019}. Radiative cooling is a sustainable, passive cooling approach that can help lessen the economic and environmental burden of cooling in our commercial sector. Radiative cooling utilizes deep-space as an infinite heat sink at a constant 3 K where thermal energy can be rejected through the highly transparent region in the atmosphere, from 8 - 13 $\mu$m, known as the ``sky-window". In order to attain a net cooling effect, materials have been developed with specifically engineered optical properties that possess the following characteristics: high reflectivity in the solar spectrum and high emissivity in the sky-window. The greater the reflectance in the solar region and the greater the emittance in the sky-window, the more cooling power a material will demonstrate.


Decades of research exploring materials that can demonstrate this exciting alternative energy phenomena has been done \cite{Harrison1978,Zeyghami2018}. More recently, photonic crystals and particle laden polymers with metallic bilayers have demonstated cooling over a full 24-hr period \cite{Rephaeli2013, Raman2014, Yang2018, Kecebas2017}. Our group has recently demonstrated full-daytime, sub-ambient radiative cooling with a proprietary high concentration BaSO$_4$-acrylic paint, as described in a non-provisional patent application (PCT/US2019/054566) filed on October 3, 2019 and published on April 9, 2020. \cite{Li2020}. 

With the recent advances of radiative cooling materials, research is now moving toward active cooling device integration into systems, as well as, passive cooling hybridization in building roofing. Goldstein et. al. bonded 3M's ESR polymer coupled with a silver reflective layer to a planar cold plate to create a non-evaporative heat exchanger, and found a heat rejection rate of 70 W$\cdot$m$^{-2}$ when flowing water at a flow rate of 0.2 L$\cdot$min$^{-1}\cdot$m$^{-2}$ \cite{Goldstein2017}. Zhao et. al. used their glass-polymer hybrid film as the planar radiating surface of an insulated water cooling module and demonstrated a water temperature drop of 10.6 K with no flow rate and with multiple water cooling modules set in parallel they achieved an impressive cooling power of 607 W at a flow rate of 26.5 L$\cdot$hr$^{-1}\cdot$m$^{-2}$ \cite{Zhao2019,Aili2019}. These studies utilized a planar radiative cooling surface bonded to a cold-plate system, leading to thermal contact resistances, as well as, thermal spreading resistances. Lamba et. al. modeled a single story commercial building with a radiative cooling roof capable of achieving 100 W$\cdot$m$^{-2}$ of cooling power. They found that a 50\% roof coverage of a building in Miami in July would almost completely eliminate the cooling load; however, this work omitted the affects of humidity on the transparency of the sky-window \cite{Lamba2018}. All existing radiative cooling systems have utilized the top surfaces that face the sky, while the bottom surfaces have not contributed to cooling thus far.


In this work we propose and successfully demonstrate concentrated radiative cooling. A radiative cooling material is coupled with an mid-infrared (mid-IR) reflector to create a novel concentrated radiative cooling (CRC) device that can radiate to deep-space from the top and bottom surfaces, simultaneously. Our device design was inspired by concentrated solar collection systems; however, it serves a fundamentally different purpose \cite{Linhua2017}.   Applying our radiative cooling paint directly on the surface of a pipe and then using a mid-IR reflector trough, we can effectively use the entire pipe's surface area to radiate thermal energy to deep-space, leading to what we define as concentrated radiative cooling. Furthermore, our radiative cooling paint is applied directly to the pipe wall which negates the thermal resistances mentioned previously, and no thermal insulation is required because the entire surface is a passive radiative cooler. We experimentally demonstrate this concentration effect. To further articulate the utility of our concentrated radiative cooling device, we predict the efficiency gain, in terms of electricity savings, that can be achieved by implementing our concentrated radiative cooling (CRC) device as a preconditioning heat exchanger to an air conditioning system, for a single story commercial building in two locations: Reno, NV and Phoenix, AZ.

\section*{Results}
\subsection*{Device Theory and Construction}

The theoretical heat flux limit for an ideal radiative cooling surface, in the sky-window region, is $\approx$150 W$\cdot$m$^{-2}$ \cite{Huang2017,Peoples2019}. However, in this work we revisit the problem from a different point-of-view, leveraging the bottom surface of an object through a mid-IR reflector to create a greater amount of area radiating to deep-space. We coat the outer surface of the copper pipe with a radiative cooling coating, as seen in Fig. \ref{CF}. A mid-IR reflector is placed under the suspended pipe to reflect the emitted thermal energy from the bottom side of the pipe to deep-space.

A more subtle, yet pertinent, design characteristic that makes this device readily usable in a diverse application space is the nesting of the pipe inside the mid-IR reflector, shown in Fig. \ref{CF}. This nesting allows an array of radiating pipes that will not radiate to one another, as illustrated in Fig. \ref{CF_Array}. Therefore the device could be installed around high rise buildings without lose of cooling power. Furthermore, the mid-IR reflector acts as a convection shield to help mitigate parasitic thermal losses from the pipe to the ambient air/wind. Another major benefit of our design is the ease of adaptation, as retrofitting it onto pre-existing systems could be accomplished with minor effort or modifications. The working fluid can flow directly through the coated pipes  negating any contact resistances or spreading resistances of the cold plate design. 

The anticipated performance of the concentrated radiative cooling system is analyzed. The schematics of the pipes without and with the mid-IR reflectors are shown in Fig. \ref{ThermResist}, along with the associated radiative transfer networks for two-surface and three-surface enclosures, respectively. For the desired optical properties needed for this device, we coated a pipe with our radiative cooling paint \cite{Li2020}; and used an annealed aluminum trough as the mid-IR reflector. The spectral properties of the radiative cooling paint and the annealed aluminum mid-IR reflector can be seen in Fig. \ref{Spectral}. The radiative cooling paint has 96\% total solar reflectance and the annealed aluminum has a reflectance of 95-99\% in the sky-window. Though the annealed aluminum trough is highly reflective in the mid-IR, the ideal mid-IR reflector would be transparent in the solar spectrum; therefore, it would not reflect solar irradiation onto the bottom surface. Since the reflectance of the annealed aluminum is $>$ 95\% it can be assumed to be a re-radiating surface in the mid-IR. 

To quantify the utility of using a mid-IR reflector, we can treat this system as a three surface enclosure with a re-radiating surface and then calculate the view factors of the pipe, illustrated in Fig. \ref{ThermResist}. The view factor is shown in Fig. \ref{ViewFactor}  as a function of the reflector opening width. Due to the pipe being nested in the reflector the pipe-to-reflector view factor ($F_{p-r}$) will always be $>0.5$. Since the exterior of the pipe cannot interact with itself, the pipe-to-sky view factor is $1 - F_{p-r}$. Without the mid-IR reflector, the ideal case would be the pipe-to-sky view factor would always be $> 0.5$, as the pipe will be blocked by its neighbors, as shown in Fig \ref{ThermResist}. With the knowledge of the view factors both with and without the mid-IR reflector, the heat flux from each scenario can be calculated. An effective way to visualize the benefit of the mid-IR reflector is to show a ratio of the heat fluxes with and without the mid-IR reflector while conserving the pipe spacing.  In Fig. \ref{ViewFactor} the ratio of the heat fluxes of the pipe with the mid-IR reflector and without the reflector is shown as a function of the mid-IR reflector opening width. As the opening gets larger the ratio will approach 2 because twice the amount of surface area radiates to deep-space. The dotted line on the figure illustrates the current design point that was used in this work, at a reflector opening of 175 mm.

\subsection*{Experimental Demonstration of Concentration}

A device was fabricated to demonstrate the concentration, shown in Fig. 3a. Monitoring the temperature of two identically coated pipes, in parallel, will provide experimental verification that our CRC device radiates with more surface area than a standalone pipe. In Fig. 3b, the pipe wall temperature of the CRC device is much lower than that of the standalone pipe; which corroborates the concentration effect. The experiments were done in West Lafayette, IN on July, 11th-12th 2019. 

Both temperature profiles show below-ambient cooling; however, the temperature drop of the coated pipe with the mid-IR reflector is approximately twice that of the standalone coated pipe. The shaded areas in Fig. 3b represent the $\pm$0.5 $^{\circ}$C error bars of the measurement. Figure 3c shows the cooling power amplification factor, which we define as the ratio of the below ambient temperature differences. The amplification factor is greater than that predicted by the three-surface enclosure model, most likely due to the testing location; the view factor of the standalone pipe to deep-space was decreased due to near-by  high-rise buildings and tall trees, while the pipe nested inside the mid-IR reflector was not hindered by its surroundings. This further illustrates the benefits of the nesting of the pipe inside the mid-IR reflector, as most real-world applications will not have a perfectly clear line-of-sight of the sky. A brief analysis is presented in the Supplemental Information that justifies how the cooling performance of the CRC device would be $>$ 2 times that of the RC device.

\subsection*{Hybrid Concentrated Radiative Cooling Air Conditioning System}

To articulate the utility of our CRC device, a building energy model was developed and employed to predict the theoretical performance enhancement, in terms of electricity energy savings from operating an air conditioning (AC) system. For this work, the CRC device was implemented as a preconditioning heat exchanger for the return air in the HVAC system, as seen in Fig. \ref{ACSys}. The preconditioning heat exchanger, referred to as the CRC system from here onward, is comprised of a water loop that flows through the CRC device where thermal energy is rejected through radiant heat transfer to deep space. The chilled water is then used in a heat exchanger to pre-cool the return air prior to the air conditioner. For comparison, the seasonal energy model simulations were also performed with the standalone pipe without the reflector, and this system will be referred to as the RC system from here onward. A temperature sensor is placed after the pre-cooling coil of the CRC system that will be connected to the thermostat of the traditional AC system. The set-point of the sensor is 25  $^\circ$C. The AC system will be turned on if the temperature of the air which travels through the heat exchanger is higher than the set-point and AC system will be turned off if the temperature drops below the set-point. In this fashion, the CRC system will be allowed to carry as much of the space cooling load as possible, and anything that remains will be handled by the traditional air conditioning system.

TRNSYS, a transient system simulation software package, was selected for performing the building energy model simulations \cite{Klein2017}. The building model was constructed based on the small commercial reference building published by the U.S. Department of Energy (DOE) \cite{Deru2011}. The weather data utilized in the simulation was the typical meteorological year (TMY) data of Reno, NV and Phoenix, AZ, respectively. The simulation time ran from May 2nd to September 31st to study the performance of the system during the summer season \cite{Wilcox2008}. The roof area of the building was 600 m$^2$. The roof coverage area of the CRC system varied, ranging from 0\% to 100\% in 10\% increments.

The energy consumption reduction of the air conditioner has a nonlinear relationship with roof coverage of the CRC system. In Fig. \ref{EConsume} the electricity usage from the AC system is shown by the black solid line as the control case, the red dashed line represents the AC electricity usage of the RC system, and the blue dotted line shows the CRC system AC electricity usage. Reno does not have an extremely high cooling demand throughout a typical cooling season, so both radiative cooling scenarios show potential benefits. However, Phoenix is much hotter, and has a greater amount of active cooling hours for TMY data; therefore, our CRC system shows greater benefits than the RC system. In Reno, the CRC system supplied more cooling capacity than was required within the building, based on the set-point temperature, i.e. the return air temperature after the CRC system was much lower than the required set-point. This led to excess or wasted cooling capacity, and ideally this cooling energy could be utilized more effectively for other subsystems throughout the building or stored for use when the heating demand increases or when the sky-window is obscured by heavy cloud coverage. Phoenix has much higher cooling loads so there is a more effective utilization of the cooling capacity supplied by the CRC system. 

The energy consumption of the air conditioner is directly related to the hours it must operate to meet the load that was not covered by the CRC system. Figures \ref{ESaveReno} and \ref{ESavePhx} show that the accumulated hours when the return air temperature after the CRC system was lower than the required set-point (hence the air conditioner is off) for Reno, NV and Phoenix, AZ, respectively. The trend of the energy saving plots is similar to the trend of the accumulative off hours, found in Figs. \ref{ESaveReno} and \ref{ESavePhx}. Thus, the major energy savings come from the reduction of the air conditioner operating hours. The CRC system saved  $> 75\%$ of the air conditioner energy consumption at 100\% roof coverage for both locations.


\section*{Discussion}

We have successfully demonstrated the concept of concentrated radiative cooling. Using a mid-IR reflector, heat transfer through radiation is enabled from the bottom surface of an object to deep-space. The nested pipe design also blocks two radiating surfaces from exchanging with one-another and provides some convective shielding. Outdoor experiments have shown that the mid-IR reflector does indeed reflect the thermal radiation from the bottom surface to deep-space, quantifiable by the temperature drop of the two pipes. Lastly, the building energy models show the theoretical electricity savings  of $>80\%$ in Reno, NV and $>75\%$ in Phoenix, AZ that can be obtained by incorporating the CRC system into an HVAC system for a single story commercial building at 100\% roof coverage.

Our proposed CRC system is not limited to the air conditioning sector; another major use for this system could be cooling power electronics in data centers. More than 50\% of the electricity utilized in data centers is used by the IT equipment and around 40\% by the cooling system. The total data center electricity consumption was about 2-2.5\% of the electricity worldwide in 2019 and expected to be around 8-9\% in the next decade \cite{Malmodin2018,Andrae2017}. Today, about 3 kW of the total power within the server's power consumption is generated as waste heat, which has been dramatically increasing due to an increase of high power electronic components \cite{Yuksel2019}. Thus, novel thermal management strategies, such as our CRC system, can be investigated and developed to create the next generation of sustainable and energy-efficient data centers.

Furthermore, the concept of utilizing the bottom surface of an object as a radiative cooler is not limited to pipe-trough designs. For instance, a planar solar cell could be cooled on the back side by coating the back with a radiative cooling material and then using a mid-IR reflector to transfer the thermal energy to deep-space. We expect this technology will be utilized in many areas to further increase utilization of passive radiative cooling. Our approach enhances radiative cooling performance while lessening the barrier to commercialization.


\section*{Methods}

\subsection*{View Factor Calculations}

In order to understand the theoretical enhancement of our mid-IR reflector, the view factors must be calculated. First we calculated the sky-to-pipes view factor ($F_{s-ps}$) for an array of pipes with no reflector, as seen in the top case of Fig. \ref{ThermResist}, using:

\begin{equation}
F_{s-ps} = 1 - \left[1 - \left(\frac{D}{s}\right)^2\right]^{1/2} + \frac{D}{s}  \times tan^{-1}\left(\frac{s^2 - D^2}{D^2}\right)^{1/2}
\label{eq:P-S-noR}
\end{equation}
where $D$ is the pipe diameter, and $s$ is the spacing between the centers of the pipe in the array \cite{Cengel2011}. Then using the reciprocity relationship, $A_iF_{i-j} = A_jF_{j-i}$, and approximating the area of the sky as $2s$ we can find the pipe-to-sky view factor $F_{p-s}$. This view factor is shown as the solid red line on Fig. \ref{ViewFactor} as a function of reflector opening which is equivalent to pipe spacing, $s$.

To find the view factors for the nested pipe and mid-IR reflector case, we treat the system as a three surface enclosure. The top surface represents the sky, as shown in the bottom case of Fig. \ref{ThermResist}. First, we calculate the pipe-to-sky view factor $F_{p-s}$ using:

\begin{equation}
F_{s-p} = \frac{1}{\pi} \times tan^{-1}\left(\frac{\nu}{h}\right)
\label{eq:P-S}
\end{equation}
where $\nu = W/(2R)$, $W$ is the reflector opening, $R$ is the radius of the pipe, and $h$ is the distance between the pipe and the sky \cite{Howell2016}.
From here all other view factors can be determined. The pipe-to-sky and pipe-to-reflector view factors are plotted as a function of reflector opening in Fig. \ref{ViewFactor}.

\subsection*{Field Experiments}

We used two copper pipes, 6.35 cm in diameter and 20 cm in length, coated with our radiative cooling paint; one with the reflector and one without, as seen in Fig. 3a, the reflector opening is 17.5 cm wide. Both setups were surrounded by a convection shield to negate the wind effects. We monitored the temperatures of the pipe wall using type-K thermocouples from the inside, three thermocouples for each pipe to assure accuracy. Two thermocouples inside the pipe were on the top surface and one was on the bottom surface. All three thermocouples attached to the pipe wall reported the same temperature within $\pm$ 0.35$^\circ$C. This level of uniformity is understandable given that the cooling powers of the top and bottom surfaces are not dramatically different, and the thermal conductivity of the copper pipe dominates the heat capacity of the combined coating and pipe system. Three thermocouples monitor the ambient temperature: one inside the convection shield of each pipe, and one outside the convection shields. All the ambient thermocouples reported temperatures within $\pm$ 0.42$^\circ$C. The data was logged with a Graphtec GL840 datalogger. 

\subsection*{Building Energy Model}

The single story commercial building utilized in this work was a reference building provided by the DOE \cite{Deru2011}. All details of this building's materials, structures, and energy efficiency ratings can be found online; however, we will quickly go over some of the major details for clarity. A door is located on the south wall, with dimension of 2.13 m x 1.8 m (H x L). There are six windows on the south/north walls and four windows on the east/west walls; with dimensions of 1.52 m x 1.83 m (H x L). The building model is cut into five simulation zones. Four zones on the perimeter of the building and one zone in the center. The exterior walls are set as the mass wall type, which has continuous insulation. The U-value of the exterior walls is 0.51 W$\cdot$m$^{-2}\cdot$K$^{-1}$; the U-value of the exterior roof is 0.316  W$\cdot$m$^{-2}\cdot$K$^{-1}$. These settings satisfy the requirements listed in the U.S. DOE commercial reference building models code \cite{Deru2011}. 

The HVAC system is a constant air volume (CAV) design, and the airflow rate is 2367 kg$\cdot$hr$^{-1}$. The water flow rate is modulated based on the cooling power from the RC system or the CRC system and cooling demands of the spaces. The RC/CRC system is installed on the top of the roof; hence, the effective roof solar absorptance decreases as the area covered by RC/CRC system increases. The solar absorptance of the roof is 0.6 and the solar absorptance of our radiative cooling paint is 0.025. The average solar absorptance of the roofis a ratio according to the covered area of the RC/CRC system which was implemented in the building model. 

The cooling power of the CRC and RC systems are calculated based on the view factor calculations and the experimentally measured cooling power of the BaSO$_4$ paint \cite{Li2020}, and can be seen below: 

\begin{equation}
q^{''}_{CRC} = 0.715 \times (1.6425 \times 10^{-8} \times T_{amb}^4 - 0.038 \times G)
\label{q_crc}
\end{equation}
\begin{equation}
q^{''}_{RC} = 0.466 \times (1.6425 \times 10^{-8} \times T_{amb}^4 - 0.038 \times G)
\label{q_rc}
\end{equation}
where $G$ is the solar irradiation. The cooling powers are modulated based on total sky cloud coverage from the TMY3 data \cite{Wilcox2008}. The predicted cooling power at 70\% roof coverage can be seen in Fig. \ref{SI_1}.

For simulating internal heat gains, a working schedule was set based on the U.S. Department of Energy Commercial Reference Building Models of the National Building Stock \cite{Deru2011}. The operating hours of the building started from 8:00 AM to 6:00 PM every weekday (Monday to Friday). There were twenty-five people who generate 100 W heat (60 W sensible and 40 W latent) individually; there were twenty-five 50 W computers; the lighting gain was assumed 5 W$\cdot$m$^{-2}$ in the building. The above internal heat gains contributed energy to the space during operating hours; moreover, the schedule of the air conditioner was the same as the operating hours.


\bibliography{Citations}

\bibliographystyle{Science}


\section*{Acknowledgments}
\subsection*{Funding}
J.P., X.L., and X.R. thank the Cooling Technologies Research Center and the Center for High Performance Buildings at Purdue University for their financial support on this work. Y.H. T.H. and J.B. thank the Center for High Performance Buildings at Purdue University for their financial support on this work. All authors thank Ray W. Herrick Laboratories and The Bechtel Innovation Design Center at Purdue University for use of their facilities.  

\subsection*{Author Contributions}
J.P. and X.R. conceived the theory. J.P., X.L., D.G, and N.F. performed the device fabrication, experimental characterization, and outdoor measurements. Y.H. and T.H. modeled the building energy envelop and savings analysis. J.P. and Y.H wrote the manuscript. All authors provided comments and edits to the manuscript. 

\subsection*{Competing Interests}
J.P. and X.R. are working toward filing a provisional patent on the work disclosed in this manuscript.  



\clearpage
\section*{}
\renewcommand{\thefigure}{\arabic{figure}}
\setcounter{figure}{0}

\begin{figure}[H]

	\begin{subfigure}{0.5\textwidth}
		\centering
		\includegraphics[width=\textwidth]{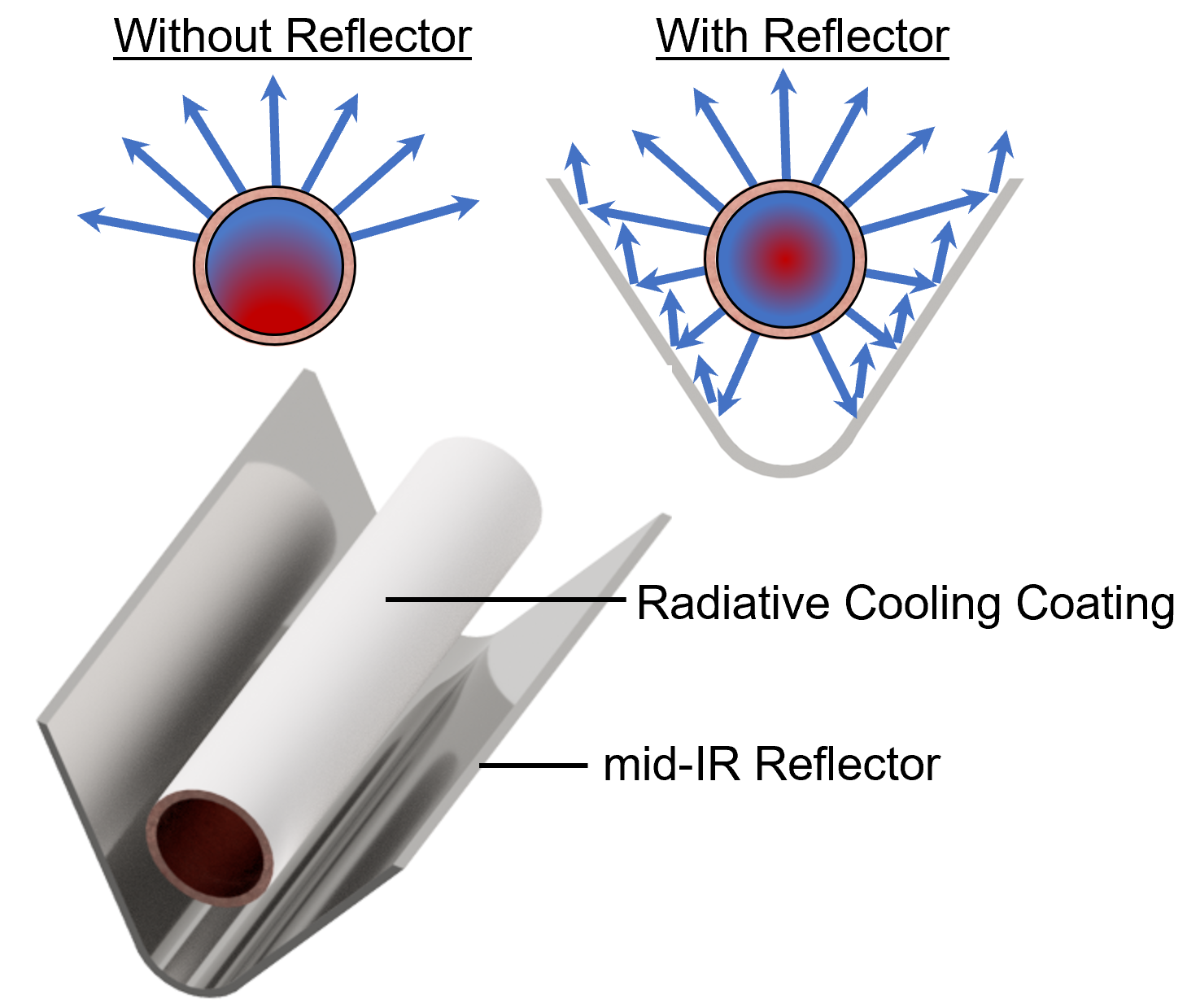}
		\caption{}
		\label{CF}
	\end{subfigure}
	\begin{subfigure}{0.5\textwidth}
		\centering
		\includegraphics[width=\textwidth]{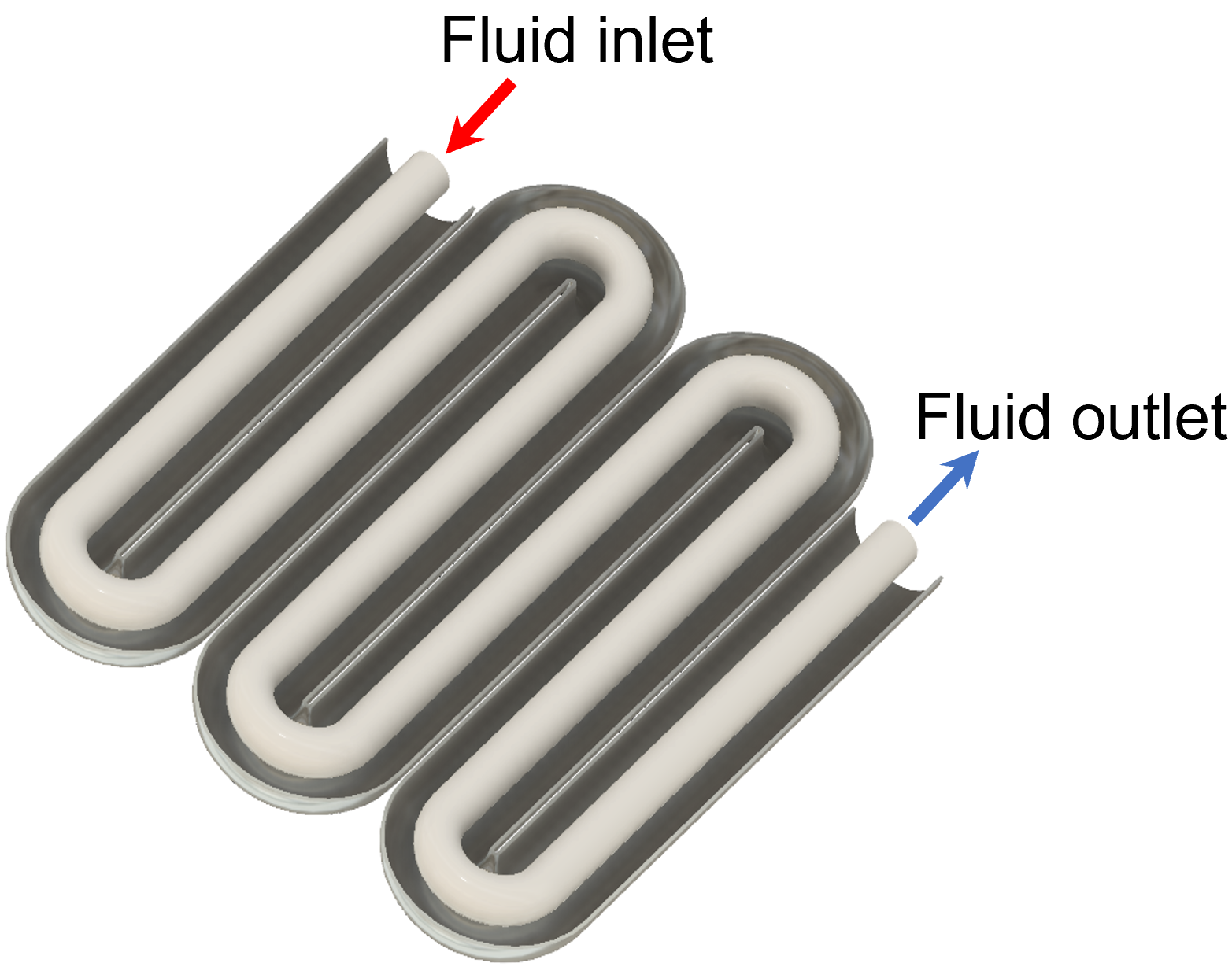}
		\caption{}
		\label{CF_Array}
	\end{subfigure}

\caption{Schematic of the ``Concentrated Radiative Cooling" device. (a) Design of our concentrated radiative cooling device and how it improves radiation from the pipe. (b) How the system would be implemented on a larger-scale with a working fluid. }
\end{figure}

\begin{figure}[H]

	\begin{subfigure}{\textwidth}
	\centering
		\includegraphics[width=0.75\textwidth]{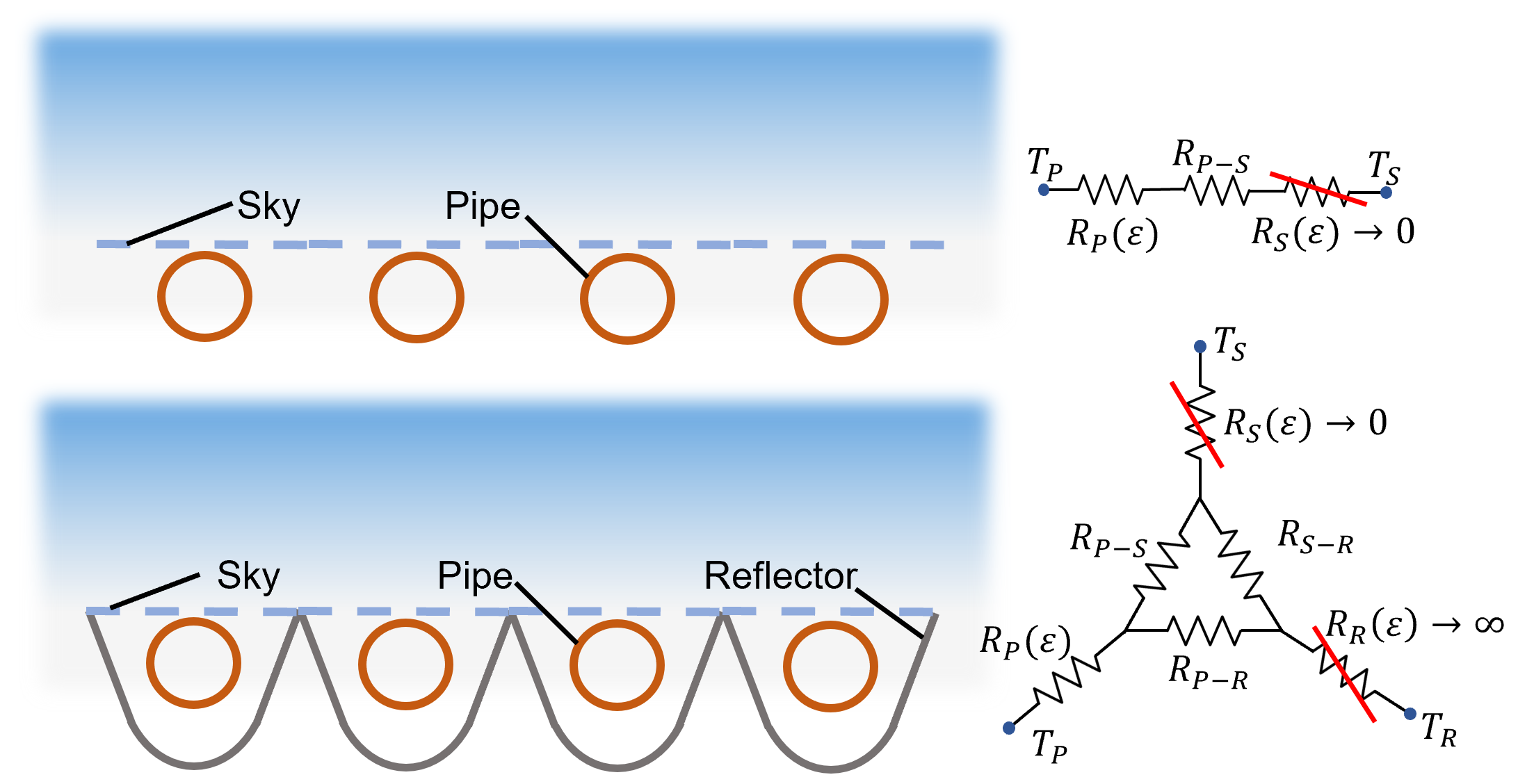}
		\caption{}
		\label{ThermResist}
	\end{subfigure}
	\vspace{1em}
	
	\begin{subfigure}{0.5\textwidth}
	\centering
		\includegraphics[width=1\linewidth]{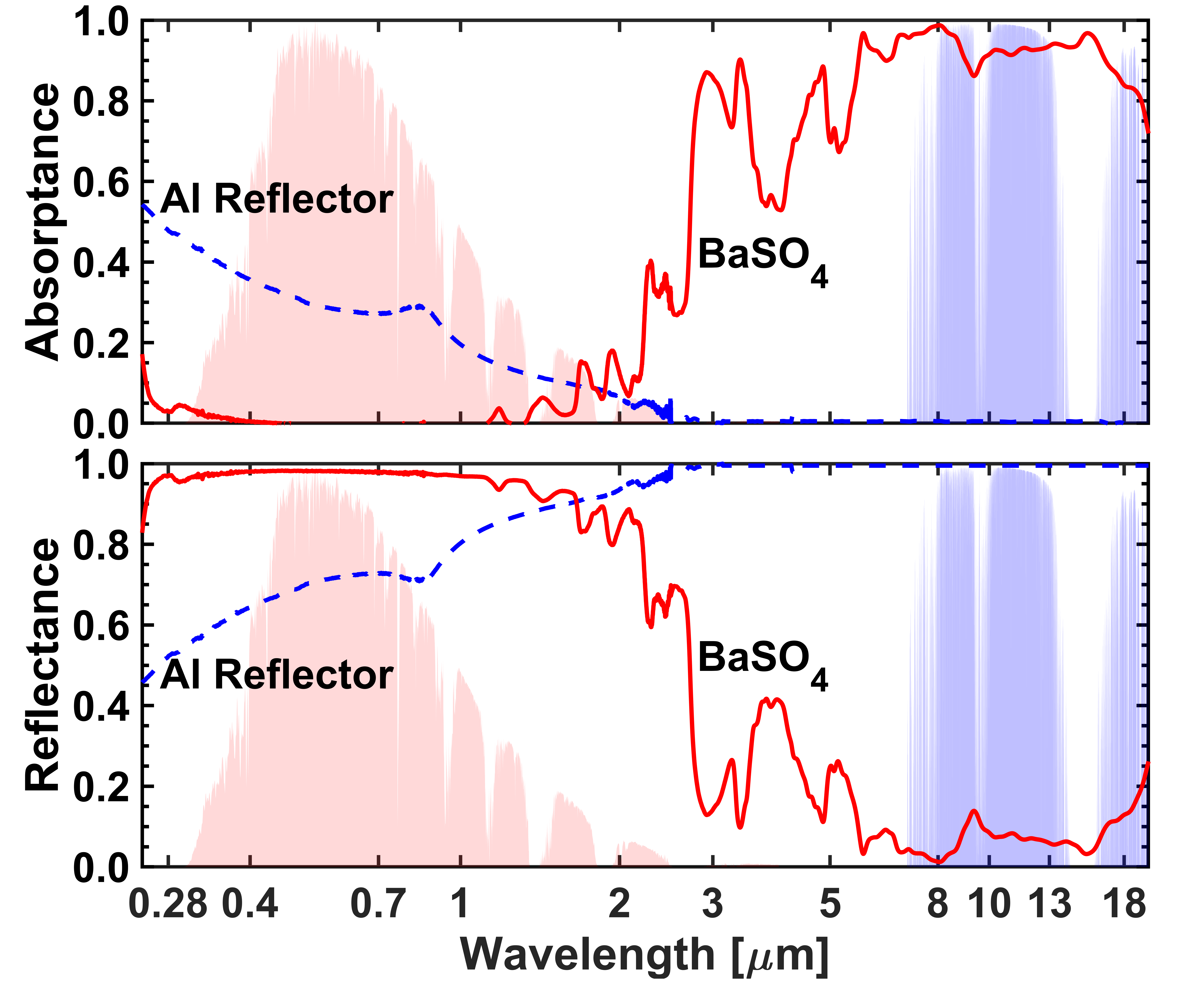}
		\caption{}
		\label{Spectral}
	\end{subfigure}	
\begin{subfigure}{0.5\textwidth}
\centering
		\includegraphics[width=\linewidth]{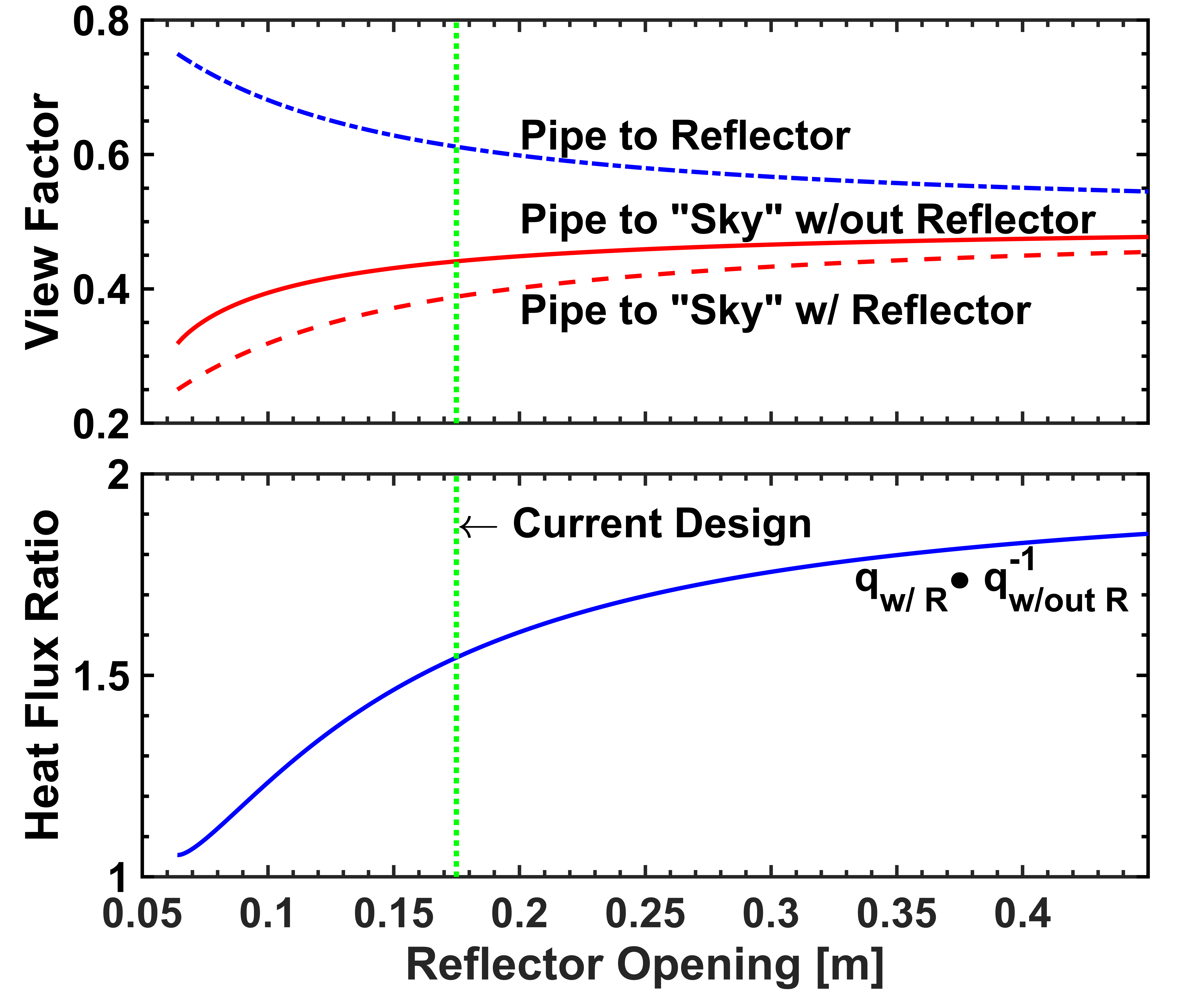}
		\caption{ }
		\label{ViewFactor}
	\end{subfigure}

\caption{Heat transfer analysis of concentrated radiative cooling device. (a) The schematic of the pipes with and without the reflector, and the associated two surface and three surface enclosures used to calculate the view factors and heat flux ratio. (b) The reflectance and absorptance of the BaSO$_4$ radiative cooling paint and the annealed aluminum for the mid-IR reflector. (c)The view factors of the pipe and the heat flux ratio of the pipe with the reflector to that without the reflector. }
\end{figure}

\begin{figure}[H]
  	\centering
	\includegraphics[width=0.93\linewidth]{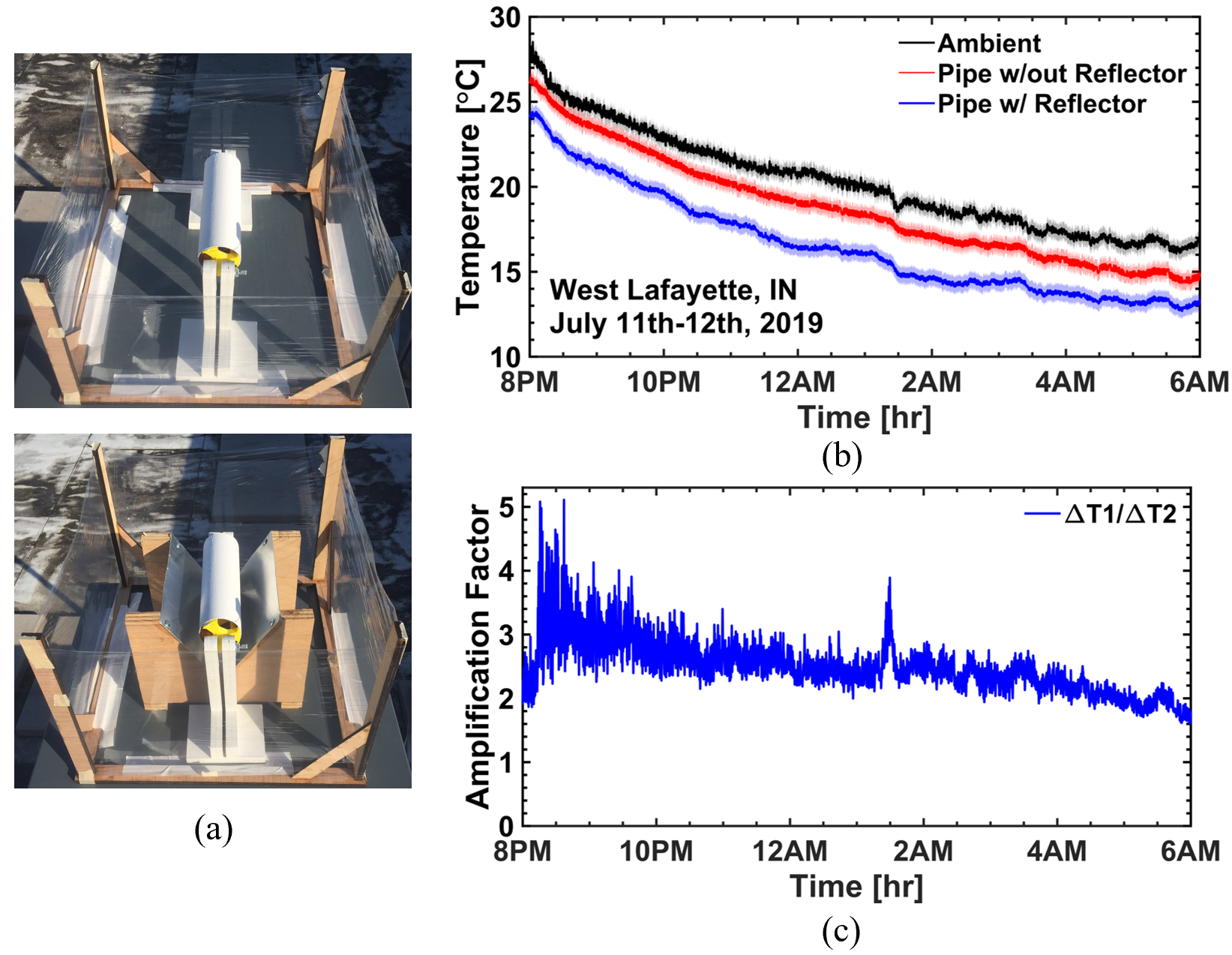}
    \caption{Field test of the concentrated radiative cooling system. (a) The outdoor setup for the pipe with and without the mid-IR reflector. (b) The temperature profiles of the ambient (black), pipe with no reflector (red), and pipe with reflector (blue). (c) The cooling power amplification factor defined as the ratio of the below ambient temperature differences between the pipe with and without the mid-IR reflector. }
    
\end{figure}

\begin{figure}[H]
	\begin{subfigure}{0.45\textwidth} 
	\centering
	\includegraphics[width=1\linewidth]{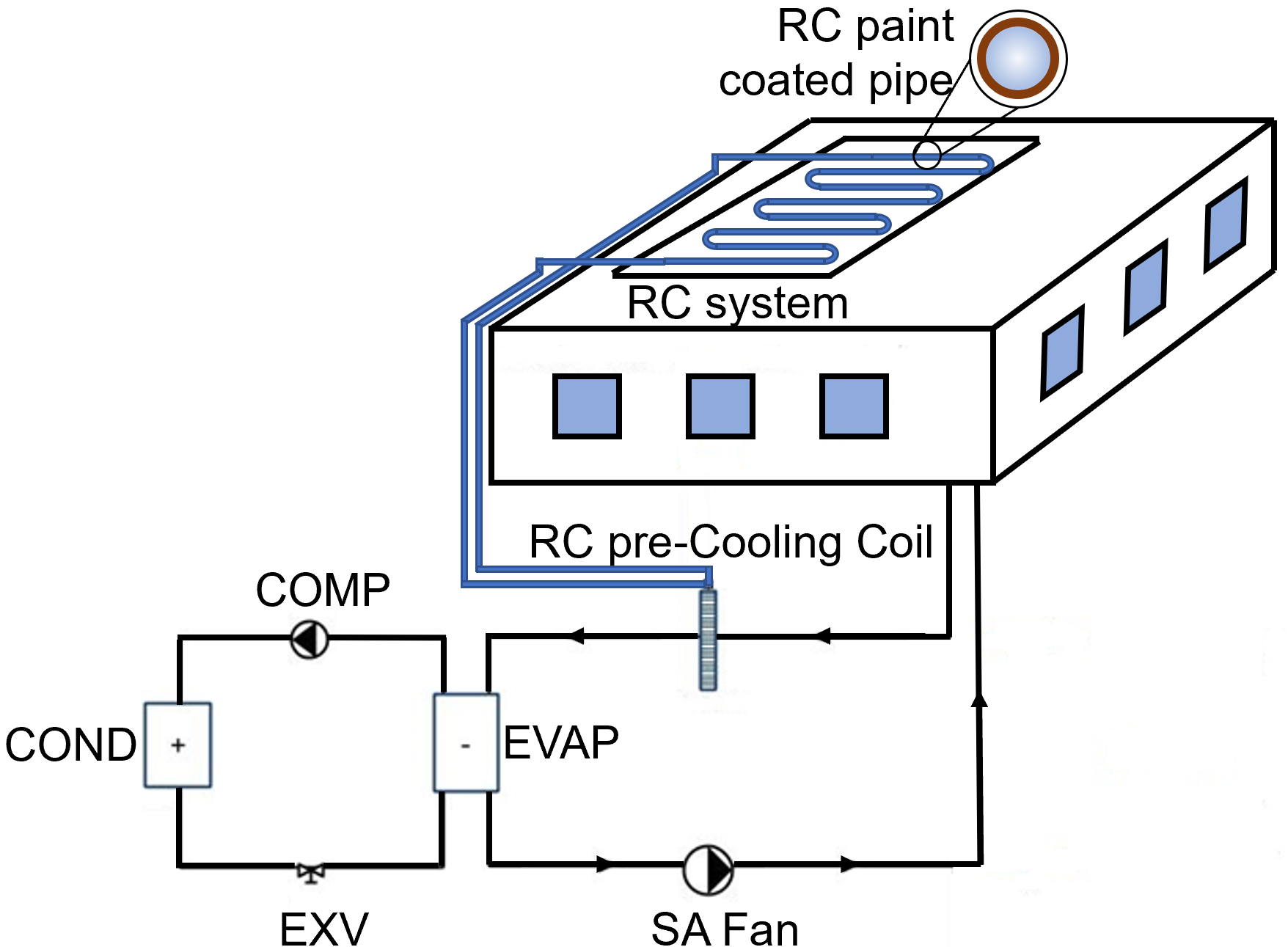}
	\caption{}
	\label{ACSys}
	\end{subfigure} 
		\hspace{1em}
	\begin{subfigure}{0.45\textwidth}
	\centering
	\includegraphics[width=1\linewidth]{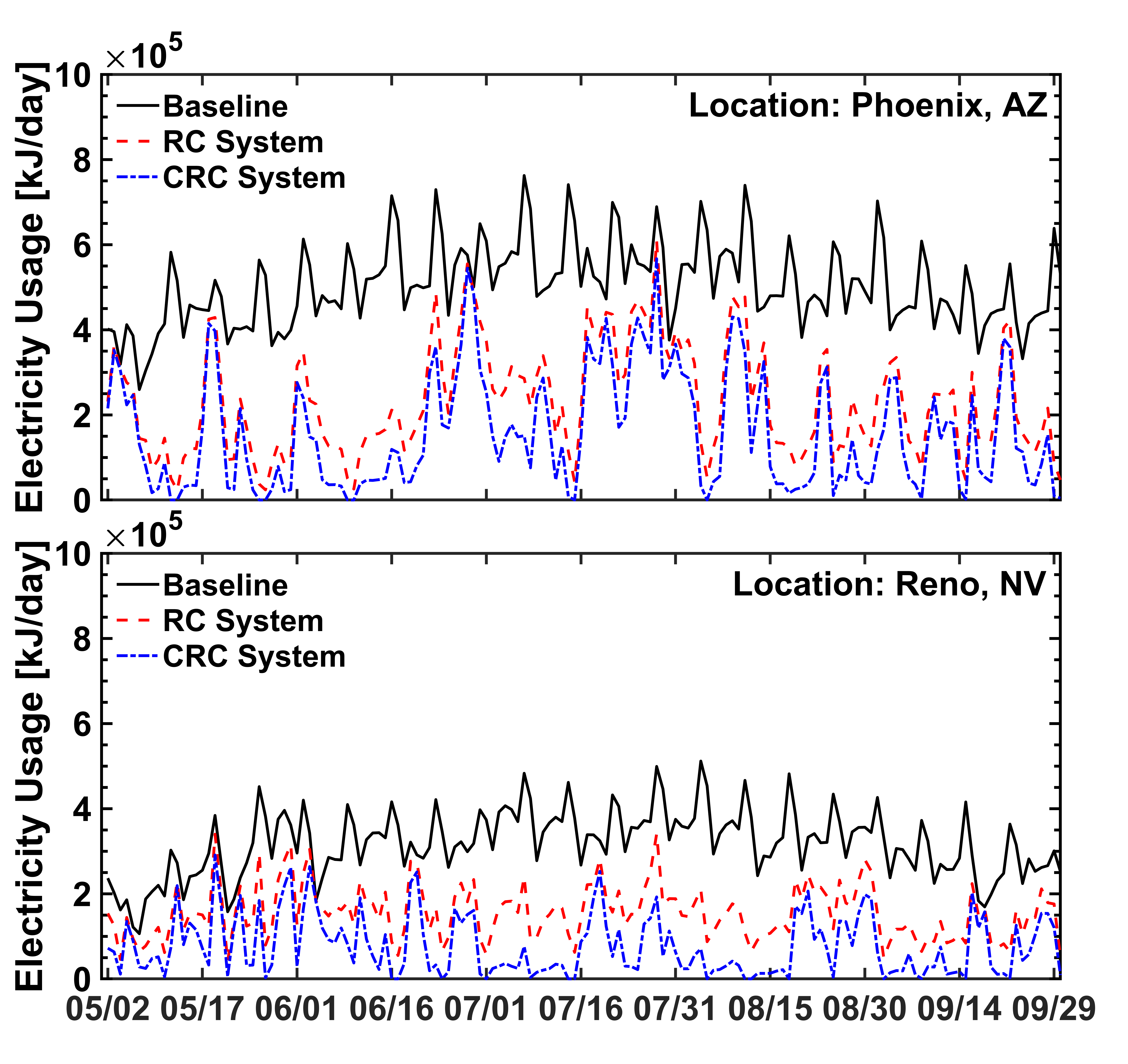}
\caption{}
\label{EConsume}
	\end{subfigure}
		
	
	\begin{subfigure}{0.45\textwidth}
	\centering
\includegraphics[width=\linewidth]{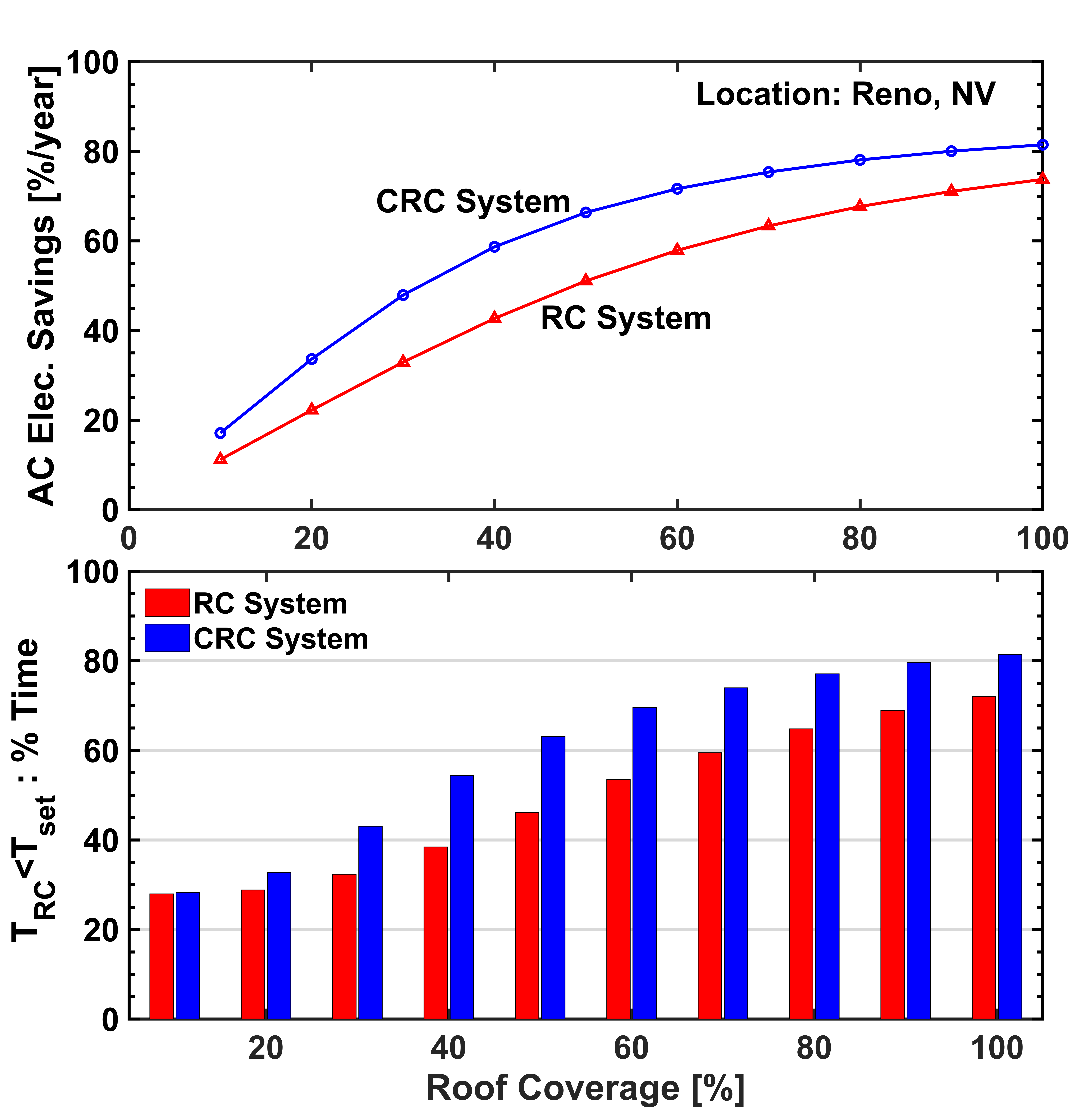}
\caption{}
\label{ESaveReno}
	\end{subfigure}	
		\hspace{1em}
	\begin{subfigure}{0.45\textwidth}
	\centering
\includegraphics[width=1\linewidth]{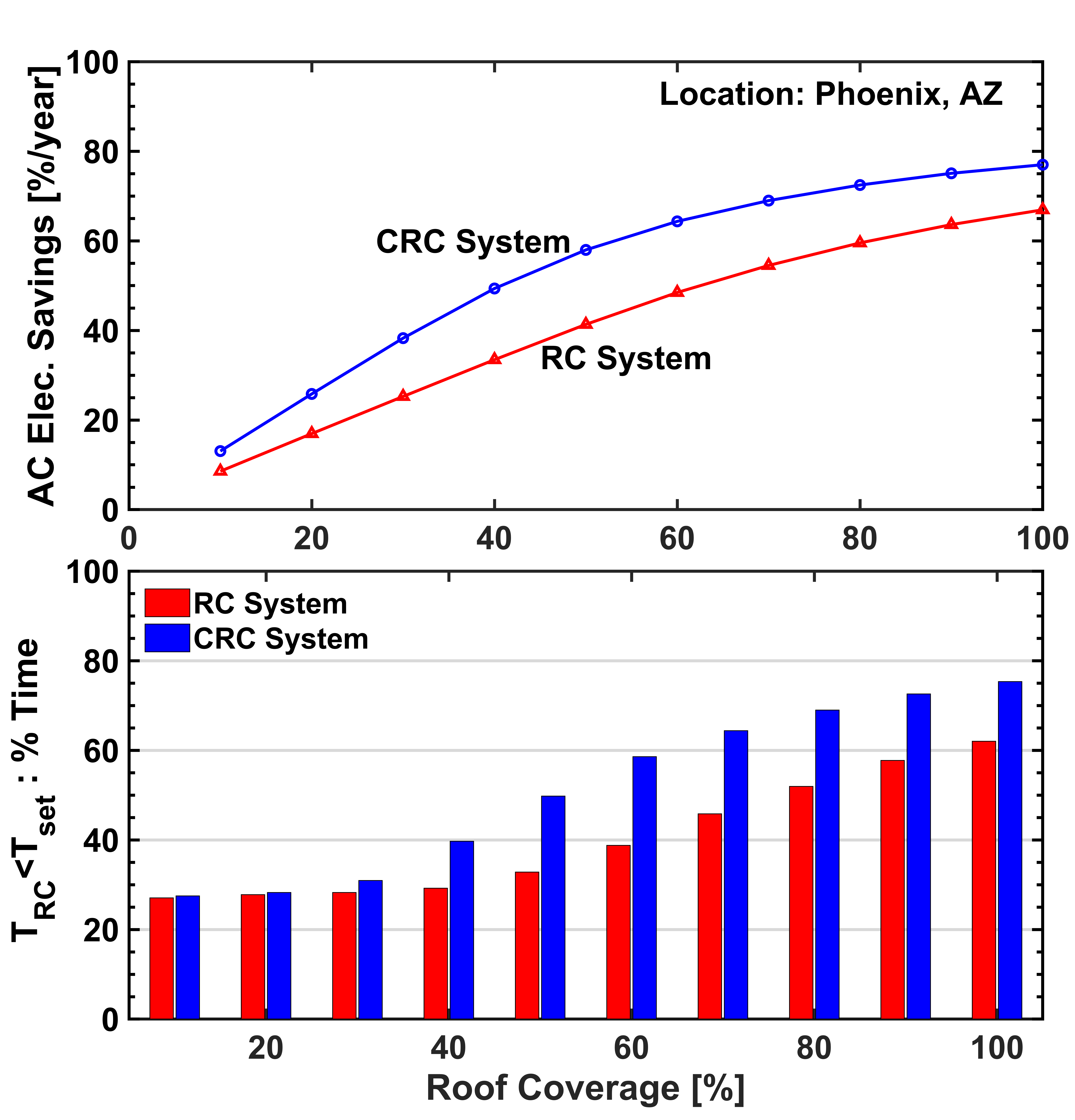}
\caption{}
\label{ESavePhx}
	\end{subfigure}

\caption{Predicted energy savings with our concentrated radiative cooling system. (a) An illustration of the cooling system used for the building simulations. (b) The electricity usage for TMY data in Reno, NV (top) and Phoenix, AZ (bottom) for three cases: the control case, RC system (standalone pipes), and the CRC system (pipes with reflectors) at 70\% roof coverage. (c) The electricity savings for Reno, NV as a function of roof coverage for both systems (top) and a bar plot to illustrate the amount of time the return air temperature after the radiative cooling systems was below the set-point temperature (bottom). (d) The electricity savings for Phoenix, AZ as a function of roof coverage for both systems (top) and a bar plot to illustrate the amount of time the return air temperature after the radiative cooling systems is below the set-point temperature (bottom).  }
\end{figure}



\section*{Supplementary materials}

\subsection*{Field Experiments}

The experimental measurements prove that the temperature drop of the CRC device is twice that of the RC device, based on the wall temperatures of the pipes. Since the experiments were conducted at night with a convection shield, we can assume that both pipes reach steady state due to two modes of heat transfer: natural convection and radiation. If we neglect the radiation to the terrestrial surroundings then we can equate the radiative cooling to the natural convection heat flux. The natural convection is considered a parasitic heat gain since the temperatures of both pipes are below ambient, due to the radiative cooling. Hence we can use the natural convection correlation for a horizontal cylinder to analyze how the temperature drop correlates to the heat flux. First, the Nusselt number correlation for a horizontal cylinder \cite{Cengel2011}:

\begin{equation}
Nu = \left[0.6 + \frac{0.387 Ra^{1/6}}{\left[1 + \left(0.559/Pr_a\right)^{9/16}\right]^{8/27}}\right]^2 
\label{eq:Nu}
\end{equation}

where $Ra$ is the Rayleigh number and $Pr_a$ is the Prandtl number. The Rayleigh number is the product of the Grashof and Prandtl number. The Grashof number has $(T_s - T_{\infty})$ in the formulation, which leads to a $\Delta T^{1/3}$ term in the Nusselt number. From there, the heat flux calculation has a $\Delta T$ term, so it can be seen that the heat flux is $\approx 2^{4/3}$ greater when the temperature drop is doubled. Using our experimental results at 11:00 PM as an example, where $T_{amb} = 20.7^\circ C$, $T_{RC} = 18.9^\circ C$, and $T_{CRC} = 16.4^\circ C$, we obtain a cooling power of 4.6 W$\cdot$m$^{-2}$ for the RC system and 10.2 W$\cdot$m$^{-2}$ for the CRC system, which is an amplification of 2.2 times. These cooling power numbers are most likely much lower than the real radiative cooling power due to the neglect of radiation to terrestrial surroundings and the humidity effects on the atmospheric transmittance.

\subsection*{Building Energy Model}

\renewcommand{\thefigure}{S\arabic{figure}}
\setcounter{figure}{0}
\begin{figure}[H]

		\centering
		\includegraphics[width=\textwidth]{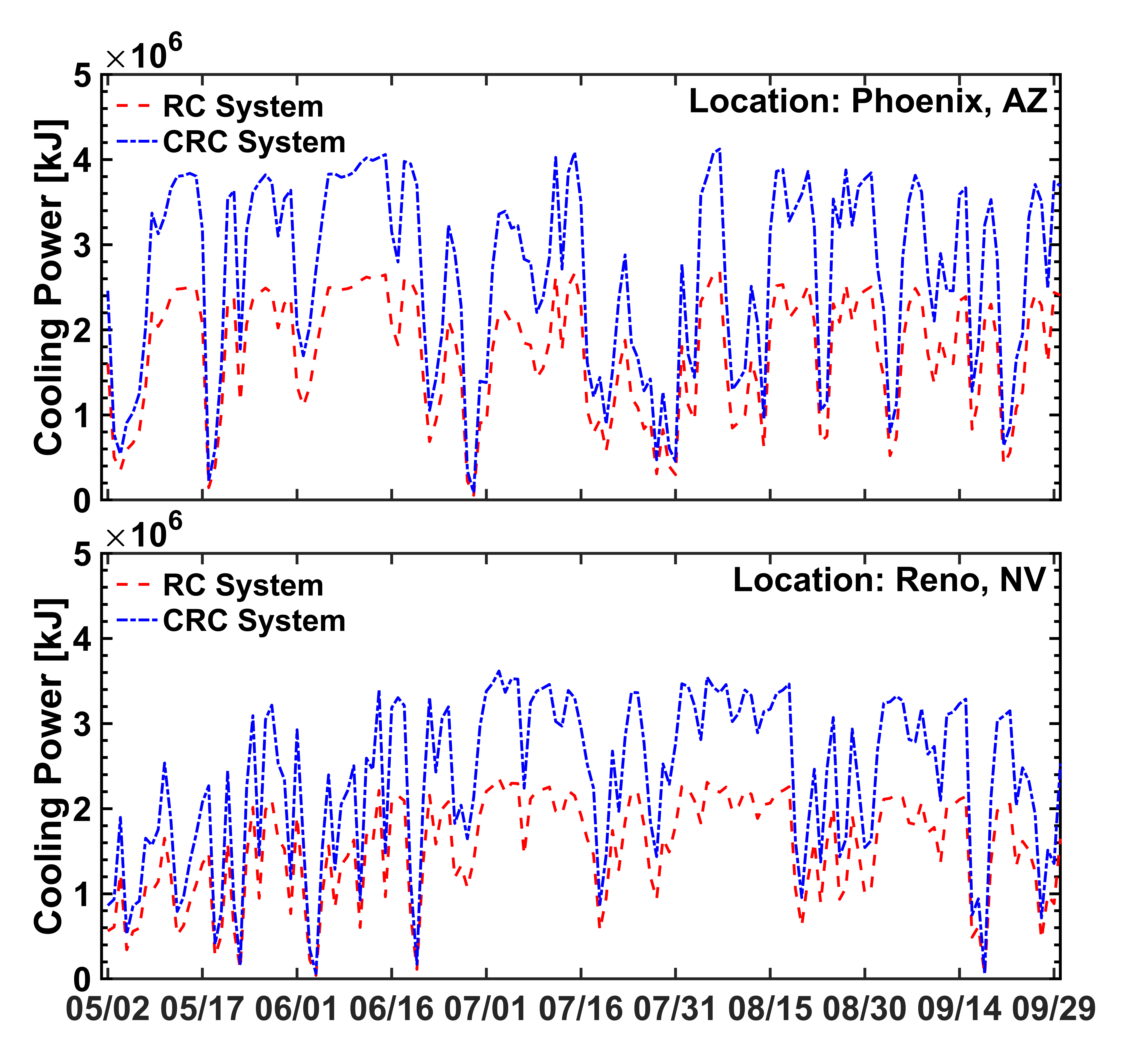}

\caption{The calculated cooling power utilized in the building analysis at 70\% roof coverage. (Top) The cooling power of the RC and CRC systems in Phoenix, AZ. (Bottom) The cooling power of the RC and CRC systems in Reno, NV. }
\label{SI_1}
\end{figure}


\end{document}